# Tailoring the Height of Ultrathin PbS Nanosheets and Their Application as Field-Effect Transistors

Thomas Bielewicz, Sedat Dogan, Christian Klinke*

*Institute of Physical Chemistry, University of Hamburg,*

*Grindelallee 117, 20146 Hamburg, Germany*

KEYWORDS: Colloidal synthesis, Colloidal nanostructures, Lead sulfide, Two-dimensional materials, Electrical transport

**Abstract**

Two-dimensional, solution-processable semiconductor materials are anticipated to be used in low-cost electronic applications, such as transistors and solar cells. Here, lead sulfide nanosheets with a lateral size of several microns are synthesized and it is shown how their height can be tuned by the variation of the ligand concentrations. As a consequence of the adjustability of the nanosheets' height between 2 to more than 20 nm charge carriers are in confinement, which has a decisive impact on their electronic properties. This is demonstrated by their use as conduction channel in a field-effect transistor. The experiments show that the performance in terms of current, On/Off ratio, and sub-threshold swing is tunable over a large range.

* Corresponding author: klinke@chemie.uni-hamburg.de

# 1. Introduction

After the discovery of graphene[1,2] various disciplines started to work on the synthesis and characterization of other freestanding two-dimensional materials,[3,4] such as $WS_2$, $MoS_2$, CdSe, or SnSe.[5-8] Compared to thin films composed of small nanoparticles, 2D nanostructures have the advantage that they do not exhibit tunnel barriers or grain boundaries in the lateral dimension, which makes them interesting for optoelectronics and photovoltaics.[9-13] Colloidal semiconductor nanomaterials have emerged as a promising route for science and industry to fabricate inexpensive and easy processable thin-film electronic devices.[14,15] Based on electronic quantum confinement in a small volume the value of their effective band gap is higher compared to bulk and a matter of size.[11,14] Lead sulfide (mineral name "galena") with its rock salt structure and with a direct bulk band gap of 0.41 eV[16] is a semiconductor material applied in infrared photo detectors[17] and solar cells.[18] As quantum confined material with a large Bohr exciton radius of about 18 nm[19,20] the optical properties of PbS can be set to the value ideal for the considered application.[21] Its capability for carrier multiplication attracted considerable attention since the effect might help to overcome the Shockley–Queisser limit and thus increase the efficiency of solar cells.[22,23]

A successful strategy to produce PbS nanoparticles is the colloidal hot-injection synthesis.[24,-26] Depending on the synthesis parameters various sizes and shapes can be obtained.[27] Despite PbS being a face centered cubic (fcc) crystal with all {100} facets being equal it has been possible to drive their growth to anisotropic materials.[28,29] Recently, we documented that two-dimensional nanosheets can be formed in the cubic system PbS.[30] The formation mechanism has been discussed and chlorinated compounds have been identified as agents altering the kinetics of nucleation and growth: Briefly, nanoparticles of about 3 nm in diameter form which then merge by *oriented attachment* to minimize the exhibited, reactive {110} facets. Through a self-assembled oleic acid monolayer on a (100) facet three-dimensional growth is hindered.

Here, we show how the colloidal synthesis of two-dimensional PbS nanosheets can be led in order to obtain different shapes and how to control their height. To understand the role of the ligands on the final morphology of the structures calculations based on the density function theory have been performed. One of the most striking features of the synthesis is that it is possible to synthesize separated, individual nanosheets with lateral sizes of several micrometers and with a tunable height between 2 and more than 20 nanometers. We demonstrate the unique possibility to tune the height and thus the size of the band gap in a field-effect transistor where the PbS nanosheets function as conduction channels. The adjustability of height and band gap allows choosing between a large On/Off ratio or a high current.

## 2. Results and Discussion

### 2.1 Synthesis without TOP

Basis of the synthesis of ultra-thin PbS nanosheets is a previously reported procedure for spherical nanoparticles,[24] which was complemented with 1,1,2-trichloroethane (TCE) to produce nanosheets.[30] Both approaches contain trioctylphosphine (TOP) as co-ligand. In the first set of experiments we omitted TOP, as its actual role is discussed rather controversially today;[31-34] even the impurities in the commercially available 97% TOP play a role in the reproducibility of nanoparticle syntheses.[35] Therefore, in the here presented studies thioacetamide (TAA), our sulfur source, was dissolved in dimethylformamide (DMF) only. TOP stabilizes the sulfur source, such that a lower effective sulfur concentration is present in the synthesis.[36] Consequently, to compensate for the missing TOP in the reaction solution, the molar ratio between lead and sulfur was increased to higher ratios by reducing the concentration of the sulfur source. In the absence of TOP less sulfur-coordinating ligand is present in the synthesis and the reactivity of the sulfur source can only be controlled at a molar ratio of 120:1 between lead and sulfur to yield nanosheets. Further, the decreased sulfur concentration leads to an almost complete consumption of the sulfur monomers at the nucleation step and ensures that only a small amount of sulfur is present at the growth step to lead the reaction path to oriented attachment rather than spherical particle growth.

To investigate the role of the ligands various OA concentrations have been used in the synthesis with the mentioned molar ratio of 120:1. The amount of the lead source lead acetate trihydrate (PbOAc), the sulfur source TAA, the co-ligand TCE, and the solvent diphenyl ether (DPE) were kept fixed. During the reaction to PbS each PbOAc molecule releases two acetate molecules. The amount of OA molecules should be at least the one of acetate stemming from the lead precursor since otherwise acetate remains in the reaction volume and functions as ligand for the nanocrystals as well, which leads to anisotropic growth.[37] With 2 mL of OA there is 1.25 times more OA than acetate in the solution which should be sufficient to convert all acetate to acetic acid. The acetic acid can then be removed completely *in vacuo*. The product of the reaction with 2 mL are mostly stripe-like rectangular nanosheets which tend to agglomerate (**Figure 1**). XRD measurements show a pronounced peak for (200) galena while other peaks are mostly suppressed (**Figure 3**). With increased amounts of OA the sheets become larger and more squared. The X-ray diffractograms for higher OA amounts show an even more pronounced (200) peak which can be explained by a texture effect: Due to their geometry the nanosheets tend to lie parallel to the substrate and only lattice planes parallel to the substrate can be measured efficiently. The (200) peak is not well defined since spherical nanoparticles are formed at the same time. The 10 mL OA sample shows even the (111) peak (**Figure 3A**) which is an indication for a certain amount of non-two-dimensional structures. With that much ligand the kinetic nanosheet product is less favored and more spherical nanoparticles can form over the course of the reaction. With more OA the lead cations are better stabilized and at later stages of the reaction lead-OA complexes are still present[32] and reactive. At the same time small PbS nanoparticles are better stabilized without forming two-dimensional structures through oriented attachement. These nanoparticles undergo spherical growth over the course of the reaction.

For the interpretation of these results we performed simulations in the frame of the density functional theory (DFT) to evaluate the ligand absorption on the different facets of PbS by employing the ORCA software.[38] The simulations **(Table 1)** show that the charged, deprotonated carboxylic acid binds much stronger to all facets than the neutral acid. The strongest adsorption energy for deprotonated acid is found on the Pb-rich {111} facets, whereas the neutral acid binds strongest to the {110} facets. Interestingly, TOP binds strongly to the S-rich {111} facets capable of stabilizing sulfur-rich sites while the negatively charged acid is much weaker. The {100} and {110} facets are strongest covered by deprotonated acid. Although TCE is a very weak ligand on all facets it is strongest on the {110} facet capable to (partially) replace the oleate due to the law of mass action and to support the growth *via* the {110} facets.

Through the very high molar ratio of 120:1 between Pb and S it is possible to synthesize anisotropic two-dimensional structures with only OA and TCE as ligands respectively co-ligands; but spherical nanoparticles are formed as well. With just 2 mL of OA the sheets are more stripe-like since most of the OA is converted to acetate which in turn stabilizes also the (110) facets. With higher amounts of OA those facets are also covered with weakly binding OA and are less stable, leading to more squared nanosheets.

**Table 1.** Adsorption energies of different ligand molecules on distinct PbS facets calculated by the DFT method. The simulations were performed using shorter versions of the ligand molecules with the functional group and a three carbon aliphatic chain (e.g. propanoic acid (PA) instead of oleic acid and tripropylphosphine (TPP) instead of trioctylphosphine). PA is the neutral acid and PA- is the deprotonated form. This does not change the qualitative conclusions.

| Adsorption energy (eV) | PbS-100 | PbS-110 | PbS-111-Pb | PbS-111-S |
| --- | --- | --- | --- | --- |
| PA | 1.116 | 1.698 | 1.074 | 0.795 |
| PA- | 3.033 | 3.793 | 4.916 | 1.748 |
| TPP | 1.414 | 1.711 | 1.751 | 3.488 |
| TCE | 0.610 | 0.882 | 0.688 | 0.616 |

## 2.2 Synthesis with TOP

In order to evaluate the influence of TOP in the synthesis we reintroduced it and varied its concentration. Those experiments show that the TOP concentration plays a minor role, but its presence improves the quality of the nanosheets synthesis. Through addition of 0.1 mL of TOP (0.2 mmol) to the precursor solution the product changes dramatically in favor for nanosheets **(Figure 2)**. Virtually, no spherical nanoparticles are formed. This must be due to the stabilization potential of TOP for sulfur resp. sulfur-rich facets and the ability to actively react with the sulfur source (to TOP-S) first before the lead precursor can and inhibit growth of PbS nuclei.[31] The stabilization process is visible through an almost doubled reaction period before the solution changes from clear to black.

Under the use of TOP it becomes obvious that with changing OA concentrations the thickness of the sheets also changes. This is not only visible by the contrast in the TEM images (**Figure 2**) and in the AFM height images **(Figure 4)** but also by the width of the (200) peaks in the X-ray diffractograms **(Figure S2)**. These peaks can be used to calculate the thickness of the sheets as no other geometrical form, like spherical nanoparticles, is interfering with this peak. Gauss fitting the (200) peak shows that the full width at half maximum (FWHM) increases with reduced amount of OA. The calculated thicknesses from the FWHM values using the Scherrer equation[39] (with a form factor of 1) had a height of 4 nm (2 mL OA), 6 nm (3.5 mL OA), 16 nm (7 mL OA), and 23 nm (10 mL OA) respectively. Measured by AFM (**Figure 4**) the nanosheets appear thinner but show the same trend (correlation co-efficient with a R² of 0.999). The sheets have a tendency to be thicker in the center and the thickness decreases continuously to the edges of the sheets. The reason for this could be that the center of a nanosheet is its "oldest" part and the longest exposed to reactive precursor. The "second layer" does not grow uniformly leaving a "canyon" type structure on top of the sheets which is still monocrystalline (**Figure S3**). It may be counterintuitive at first, that through more ligand in the solution the nanosheets become thicker but OA and oleate are not equally good ligands for lead and the respective crystal facets in lead sulfide (see DFT results). Through the synthesis parameters the amount of oleate is always restricted to double the molar ratio of the used lead acetate. At best, no acetate is present in the precursor solution at the end of the degassing procedure. With an increasing amount of OA in the synthesis also a higher amount of OA compared to oleate will be present in the synthesis volume and on the surface of the nanosheets. Considering the law of mass action, the more OA is present the higher the probability that it is binding to a crystal surface, superseding oleate. Since OA is a weak ligand it can be removed far easier from those crystal surfaces by additional monomer than oleate. The decay of the lead complex to monomer is the slower the more OA is present. Thus, at the growth step of the reaction there is more monomer present the higher the OA concentrations.[32] Consequently, with higher OA concentrations the thickness of the nanosheets is larger at otherwise unchanged reaction parameters.

By doubling the amount of TOP in the precursor solution no significant changes can be observed in the TEM. This might be due to the fact that 0.1 mL TOP corresponds to ten times the molar amount of TAA at a lead to sulfur ratio of 120:1 and another doubling does nothing for the stabilization of the sulfur source or sulfur-rich facets.

By omitting TCE from a reaction with TOP and 3.5 mL OA the products change to agglomerated nanoparticle **(Figure S4)**. The XRD supports a 3D material, as all galena peaks are present. Even at very high molar ratios of lead and sulfur, TOP and OA is not enough to enable two-dimensional oriented attachment and the chloroalkane remains a crucial component for the formation of two-dimensional PbS structures.

Due to the 2D density of states and scattering (structure is larger than the wavelength) optical absorption measurements do not show prominent excitonic features **(Figure 5)**. Thus, it is difficult to determine the exact position of the absorption edge. Anyhow, the spectra show that the absorption shifts to longer wavelength with increasing nanosheets thickness. For the nanosheets synthesized with 2 mL of OA we estimate the absorption edge to be at about 1800 nm. Using the XRD height of 4 nm in the effective-mass approximation for the calculation of the effective bandgap yields 1655 nm which is in reasonable agreement considering the use of this rough approach. For the synthesis using 3.5 mL OA (6 nm) the calculation gives an effective band gap corresponding to 2275 nm, for 7 mL (16 nm) 2966 nm, and for 10 mL (23 nm) 3023 nm which is basically the bulk value.

**2.3 Field-Effect Transistors**

In order to demonstrate the power of the synthesis we prepared field-effect transistors (FETs) made of PbS nanosheets of various thickness. Recently, we reported on PbS nanosheet field-effect transistors (FETs) which outperformed the best state-of-the-art colloidal materials in their pristine form.[40] Those individual PbS nanosheet FETs contacted by symmetrical gold-contacts showed p-type behavior and a modest transfer characteristic at room temperature whereas at lower temperatures they exhibited a pronounced p-type characteristic. At lower temperatures it became possible to completely switch off the current by the gate voltage. Here now, we show that the performance of FETs is a matter of height resp. band gap.

To explore the FET behavior a low density suspension of PbS nanosheets in toluene was spin-coated on a silicon substrate covered with 300 nm thermal oxide. The silicon is highly n-doped and is used as back gate. Individual PbS nanosheets are contacted by thermally evaporated gold electrodes *via* electron-beam lithography (**Figure S5**). For each PbS nanosheet thickness we investigated at least seven devices. **Figure 6** displays the room temperature output and transfer characteristics for three representative devices. The thicknesses of the individual PbS nanosheets were measured by AFM. Keeping in mind that the nanosheets are covered with a monolayer of OA/oleate (1.8 nm in length) on each side yields a height of the inorganic part of 2.4, 4.4, and 11.4 nm. Therefore, for the device behavior we take the AFM thicknesses and not those from the XRD as these represent the actual measured ones while the XRD represents a mean thickness. Information about the the height of the inorganic part and the effective band gap of the PbS nanosheets are displayed in **Table S1**. The effective band gap of PbS nanosheet is calculated *via* the "particle-in-a-box" approach by considering only its inorganic part. Nanosheets with a height over 11.4 nm did not show any significant confinement.

All devices show p-type behavior indicated by a decrease of the lateral current $I_{DS}$ while sweeping the gate voltage from negative to positive values. This agrees with our previously reported PbS nanosheet FETs that favor an accumulation of holes in the channel.[38] The $I_D$-$V_{DS}$ sweeps show that with decreasing nanosheet thickness the current decreases as well (**Figure 6A**). For example, the zero-bias conductance drops from $1.5 \cdot 10^{-9}$ S for the device with an 11.4 nm nanosheets, to $4.3 \cdot 10^{-11}$ S for the 4.4 nm nanosheet, and $1.5 \cdot 10^{-13}$ S for the 2.4 nm thick nanosheet. Even if the height is taken into account and the conductivity is calculated, the thicker nanosheets still conduct much better than the thinner ones. On the other hand, in the $I_D$-$V_g$ measurements the 2.4 nm sheet shows an On/Off ratio of 172.95, the 4.4 nm sheet an On/Off ratio of 6.85, and the 11.4 nm thick sheets a value of 1.75. Another key parameter to evaluate the transistor performance is the sub-threshold swing, which is defined as $S = dV_g / d(\log I_{DS})$. We calculate for the 2.4 nm nanosheet a sub-threshold swing of remarkable 3.08 V/dec. The sub-threshold swing increases with nanosheet thickness, for the 4.4 nm nanosheet we obtain $S = 5.66$ V/dec and for the 11.4 nm nanosheet $S = 17.18$ V/dec. Detailed output and transfer characteristics of all devices can be found in the Supporting Information **(Figure S6 and S7)**.

In order to understand the behavior of the transistors we consider the valence and conduction band levels of bulk PbS, the Fermi levels of the gold leads and PbS, and the confinement in the PbS nanosheets calculated by the particle-in-a-box approach with a bulk band gap of 0.41 eV and the effective masses for electron and holes of $0.12 \cdot m_e$ and $0.11 \cdot m_e$ resp.[41] In contact PbS nanosheets and gold leads form Ohmic contacts for holes at the interface and Schottky barriers for electrons (for a detailed picture of the band alignment see **Figure S8**). This leads to a pronounced current based on holes as majority carriers resulting in p-type behavior. The thinner the nanosheets are the more lies the valence band below the Fermi level of the gold leads resulting in a lower hole current. At the same time the conduction band shifts in opposite direction leading to an increased band gap and thus, larger On/Off ratios (clearer separation of electron and hole current). The increase of the band gap with decreased nanosheet thickness also leads to a reduced thermionic and tunneling current of electrons in the conduction band. Thicker nanosheets possess an almost bulk-like, small band gap and band alignment. Thus, electron and hole currents cannot be separated efficiently. Together with the thermionic and tunneling current this leads to a high current and a reduced switching. Thus, the p-type behavior can be explained by the band alignment. Similar behavior has been observed for PbS nanoparticle films.[42] Another contribution could be due to the absorption of oxygen on the PbS nanosheets as observed for PbS nanoparticles;[43] but as the surface-to-volume ratio is smaller for nanosheets compared to nanoparticle films and the nanosheets are continuous we evaluate this contribution as less significant.

The tuneability of the band gap (by nanosheets height), current, On/Off ratio, and sub-threshold swing at room temperature allows for an optimization of the material for the relevant application (high current, clear switching, or high solar cell efficiency).

## 3. Conclusion

Through the variation of just one ligand concentration the thickness of two-dimensional PbS nanostructures can be controlled precisely. The obtained nanosheets possess lateral dimensions of several microns and a height ranging from 4 to more than 20 nm (XRD values) depending on the amount of OA used in the synthesis. By introducing TOP to the synthesis the nanosheets lateral dimensions increase and virtually no spherical nanoparticles are formed whereas without TOP spherical nanoparticles can be found. The electrical transport characterizations demonstrate that thin nanosheets show steep sub-threshold swing and a high On/Off current ratio, whereas thicker nanosheets exhibit a comparatively high On current. This makes the presented PbS nanosheets an interesting, tunable and versatile material for inexpensive electronic devices like transistors or solar cells.

## 4. Experimental Section

*Synthesis*: All chemicals were used as received: lead(II) acetate tri-hydrate (Aldrich, 99.999%), thioacetamide (Sigma-Aldrich, >= 99.0%), diphenyl ether (Aldrich, 99%+), dimethyl formamide (Sigma-Aldrich, 99.8% anhydrous), oleic acid (Aldrich, 90%), trioctylphosphine (ABCR, 97%), 1,1,2-trichloroethane (Aldrich, 96%). In a typical synthesis a three neck 50 mL flask was used with a condenser, septum and thermocouple. 860 mg of lead acetate trihydrate (2.3 mmol) were dissolved in 10 mL of diphenyl ether and 2 – 10 mL of OA (5.7-28 mmol) and heated to 75 °C until the solution turned clear. Then a vacuum was applied for 3.5 h to transform the lead acetate into lead oleate and to remove the acetic acid in the same step. The solution was heated under nitrogen flow rate to the desired reaction temperature of 130 °C while at 100 °C 0.7 mL of TCE (7.5 mmol) was added under reflux to the solution and the time has been started. After 12 minutes 0.23 mL of a 0.04 g TAA (0.5 mmol) in 6.5 mL DMF was added to the reaction solution. After 5 minutes the heat source was removed and the solution was let to cool down below 60 °C which took approximately 30 min and afterwards centrifuged at 4000 rpm for 3 minutes. The precipitant was washed two times in toluene before the nanosheets were finally suspended in toluene again for storage.

For a synthesis with TOP the amounts (0.1-0.4 mL; 0.2-0.9 mmol) were added before heating of the solution to 75 °C and the vacuum application.

*TEM*: images were done on a JEOL-1011 with an acceleration voltage of 100 kV. The TEM samples were prepared by diluting the nanosheet suspension with toluene and then drop casting 10 µL of the suspension on a TEM copper grid coated with a carbon film.

*XRD:* measurements were performed on a Philips X'Pert System with a Bragg-Brentano geometry and a copper anode with a X-Ray wavelength of 0.154 nm. The samples were measured by drop-casting parts of the suspended nanosheets on a <911> or <711> grown silicon substrate.

*AFM:* measurements were performed on a Veeco Dimension 3000 AFM in contact mode. The samples were prepared by spin-coating the nanosheet suspension on a silicon wafer.

*Absorption spectra*: measurements were performed by drop casting enough material onto a quartz cuvette and wait until the solvent evaporated. The absorption measurements were done with a CARY500 UV-Vis-NIR in transmittance.

*Transport measurements*: The substrates used for the transport measurements were silicon wafer covered with a 300 nm thermal oxide layer. The Si/SiO$_2$-substrates were patterned with alignment marks by e-beam lithography, followed thermal evaporation of gold (1 nm Ti + 40 nm Au) and lift-off in acetone. In toluene suspended nanosheets were transferred on these substrates via spin-coating. Relative to the alignment marks the individual nanosheets were identified using an Olympic BX51 optical microscope. After that the individual nanosheets were contacted by e-beam lithography with gold electrodes (1 nm Ti + 40 nm Au). A sample device can be found in the Supporting Information (Figure S5).

The devices were characterized in a Lakeshore-Desert probestation under vacuum using a Keithley 4200 semiconductor parameter analyzer.

*Simulations:* For the interpretation of the results we performed simulations in the frame of the density functional theory (DFT) to evaluate the ligand absorption on the different facets of PbS by employing the ORCA software.[38] The calculations were performed with the LANL ECP[44] / def2-TZVP/J[45] basis set for Pb and the def2-TZVP/J one for the lighter elements. Furthermore, the LDA exchange functional and the correlation functional VWN-5[46] are used as DFT framework. To keep the simulation time at a feasible level, we simulated shorter versions of the ligand molecules with the functional group and a three carbon aliphatic chain (e.g. instead of oleic acid we simulated propanoic acid). This does not change the qualitative aspects of the results. The adsorption energies were calculated by comparing the sum of the separate energies of the PbS crystal and the ligand molecule with the total energy of the complete system. During simulation under aperiodic boundary conditions, the PbS crystal was fixed to the experimental lattice constant and the geometry of a pristine facet while the ligand molecule was free to relax.

*Band level calculations:* The band levels were calculated with the quantum mechanical particles-in-a-box approach: in X and Y direction there is no confinement for the nanosheets. Only the Z direction has to be calculated using the equation $E_{cond,eff} = E_{cond,bulk} - h^2/8m_e^* L_Z^2$ for the conduction band level with the effective conduction band level $E_{cond,eff}$, the bulk conduction band level $E_{cond,bulk} = 4.6$ eV,[47] the Planck constant $h$, the effective mass for electrons in the material $m_e^* = 0.12\ m_e$, and the height of the nanosheets $L_Z$. For holes a corresponding formula $E_{val,eff} = E_{val,bulk} + h^2/8m_e^* L_Z^2$ was used ($E_{val,bulk} = 5.0$ eV, $m_h^* = 0.11\ m_e$). The Schottky barriers are calculated by subtraction of the difference in work functions. The effective band gap was calculated by $\Delta E_{eff} = \Delta E_{bulk} +$ $h^2/8L_Z^2(1/m_e^*+1/m_h^*)$, where $\Delta E_{bulk}$ is the bulk band gap. Effectively, this is the difference between $E_{cond,eff}$ and $E_{val,eff}$.

**Supporting Information**

Supporting Information is available from the Wiley Online Library or from the author.

**Acknowledgments**

The authors thank the European Research Council (Seventh Framework Program FP7, Project: ERC Starting Grant 2D-SYNETRA) for funding. CK acknowledges the German Research Foundation DFG for a Heisenberg scholarship.

**Figures**

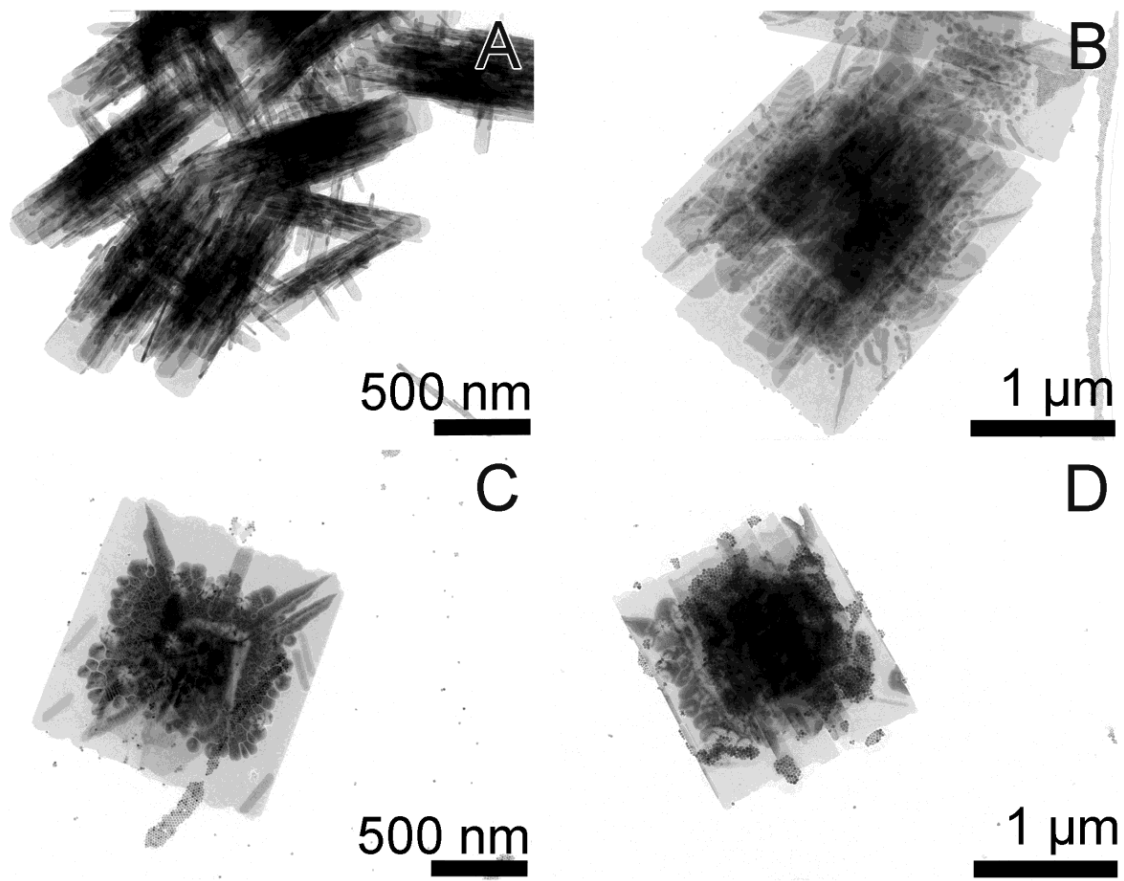

**Figure 1.** Synthesis without TOP varying the OA concentration: The OA concentration increases from 2 mL (A) to 3.5 mL (B) to 7 mL (C), and to 10 mL (D) respectively. The sheets are becoming more squared with increasing OA amounts. Spherical nanoparticles are present as well. The nanosheets show a secondary growth on top of a thinner sheet.

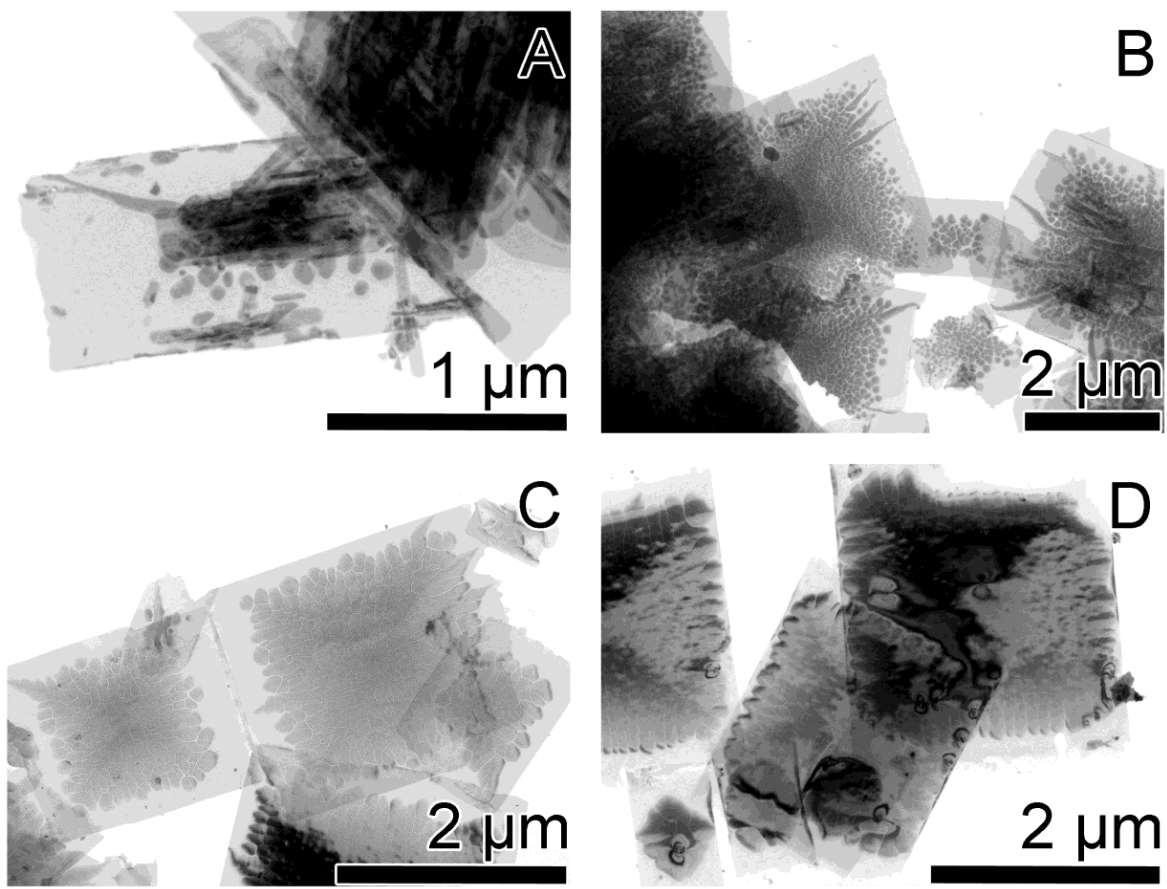

**Figure 2.** Synthesis varying the OA concentration in the presence of TOP: (A) shows the product of a reaction with 2 mL of OA. The nanosheets are very thin but a second layer which is already growing on the smooth surface is visible through darker spots, especially in the center of a nanosheet. The nanosheets are also laterally larger than in the reaction without TOP. (B) shows the product of a reaction with 3.5 mL of OA. Further vertical growth is now better visible. With 7 mL of OA, as shown in (C), the sheets seem to be much thicker on the whole surface due to the strong contrast in the TEM while the border still seems to be thinner than the rest of the sheet. (D) With even more OA (10 mL) the surface becomes smoother as if the layer is now fully grown out while the borders still remain thin.

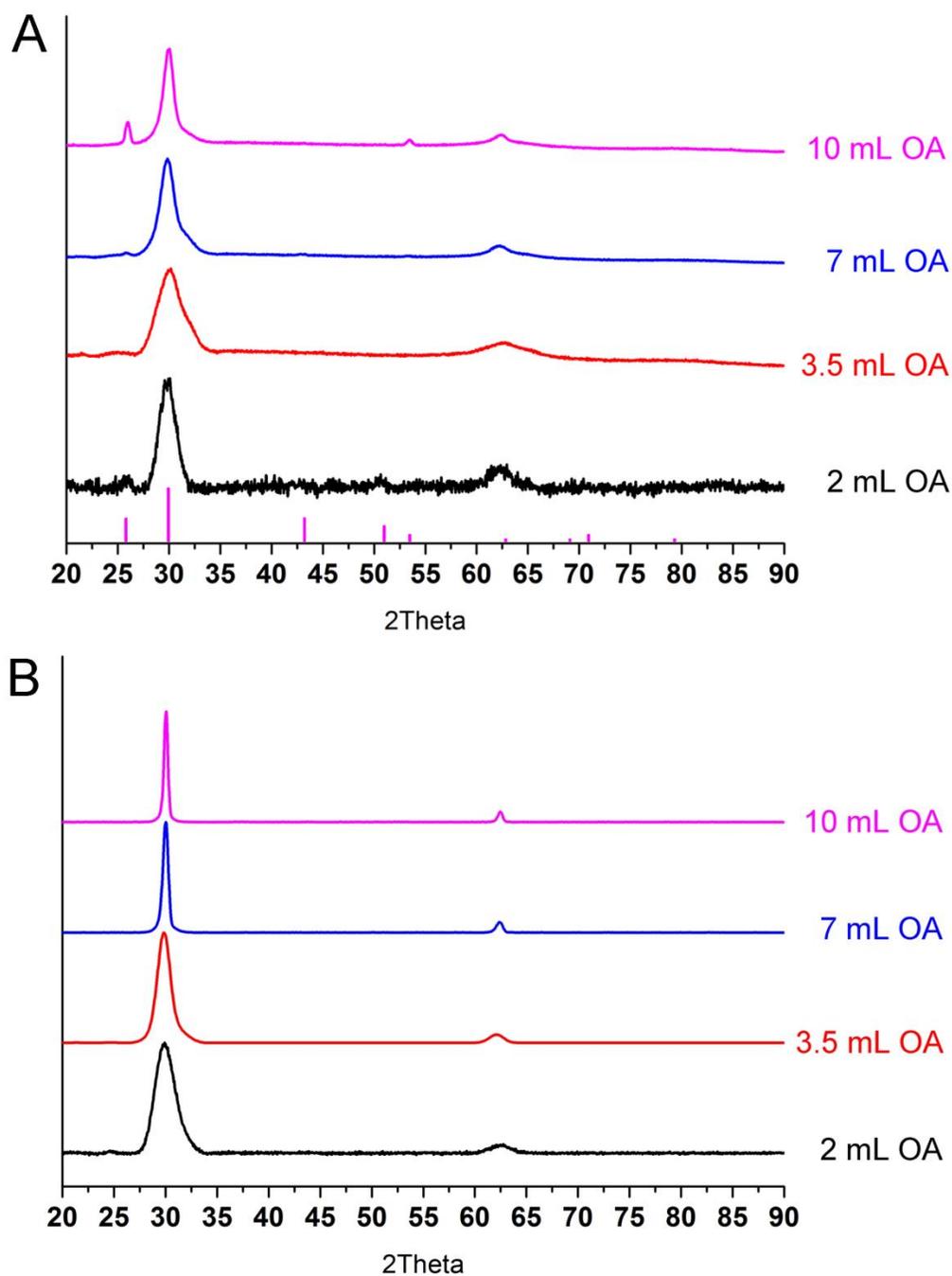

**Figure 3.** XRD of PbS nanosheets with different OA amounts in the synthesis. (A) shows the syntheses without TOP while (B) shows the diffractograms for syntheses with an addition of 0.1 mL TOP. (A) also shows galena peaks of a macroscopic PbS crystal. The diffractograms with TOP show only the peaks for the (200) crystal planes at 29.9° and the (400) crystal planes at 62.4° due to a texture effect of the two-dimensional material.

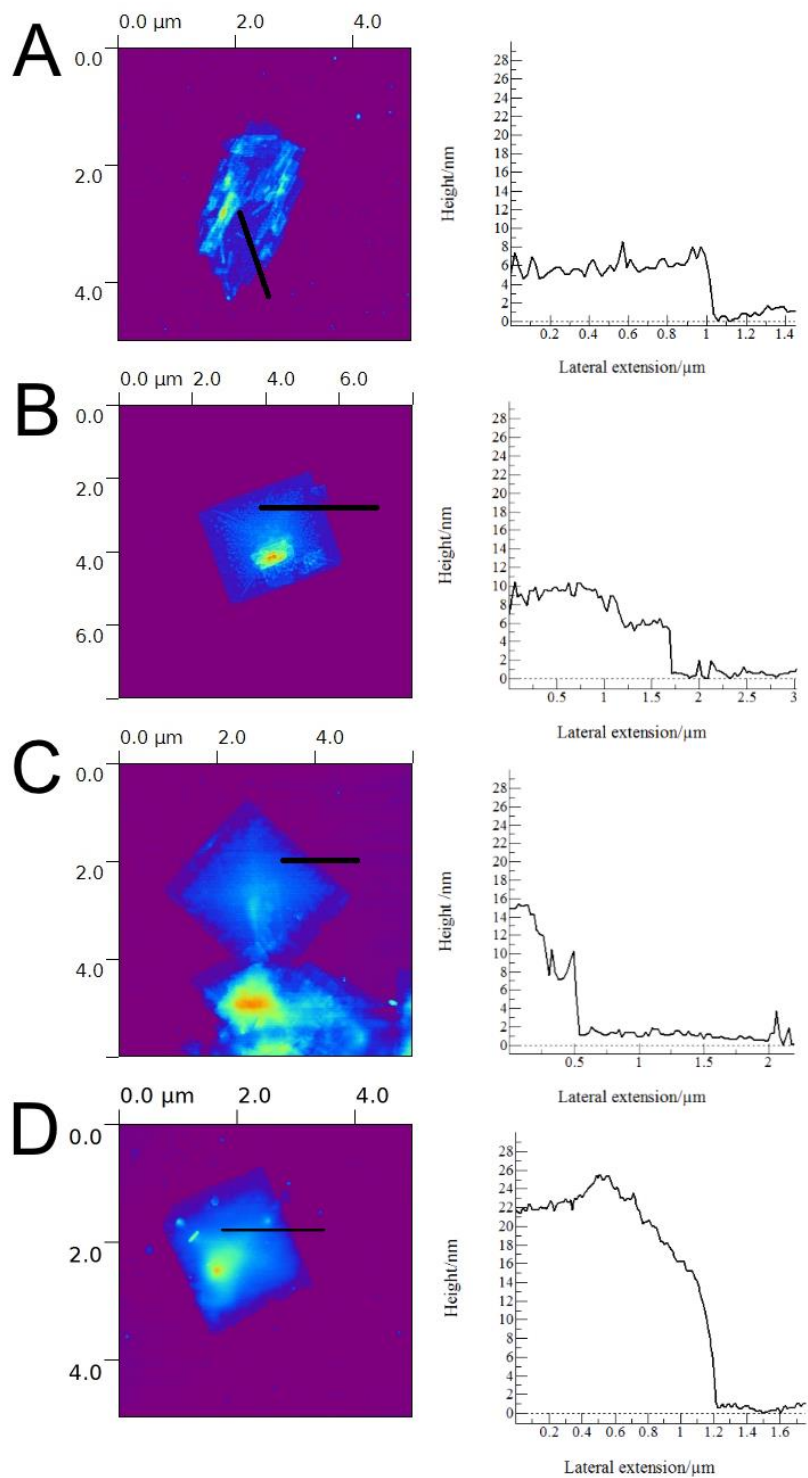

**Figure 4.** AFM images and height profiles of individual PbS nanosheets: They possess total heights of (A) 6 nm, (B) 8 nm, (C) 15 nm, and (D) 25 nm. The total heights include a top and a bottom layer of self-assembled OA of a thickness of about 1.8 nm.

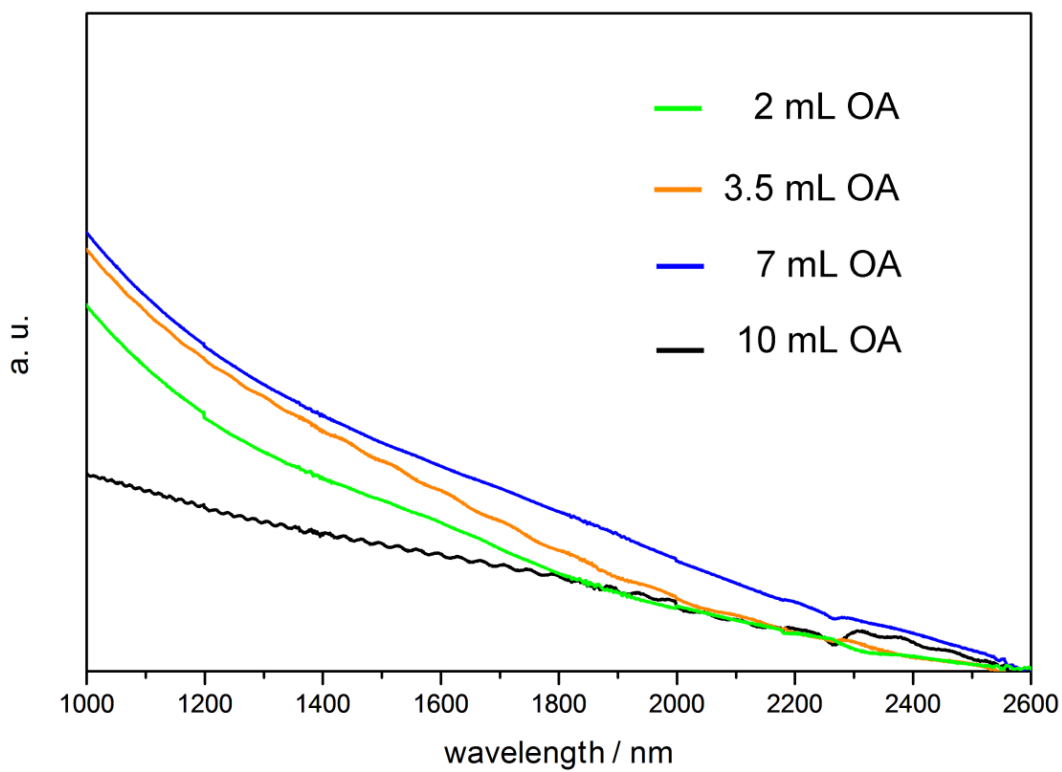

**Figure 5.** Absorption spectroscopy on PbS nanosheets of different height with their respective amount of oleic acid used in the synthesis. All spectra have been normalized at a wavelength of 2940 nm. The green line represents the 4 nm sheets, the orange line the 6 nm, the blue line 16 nm and the black line the 23 nm sheets (thickness taken from the XRD). The absorption wavelength of the thickest nanosheets lies beyond the capabilities of the detector.

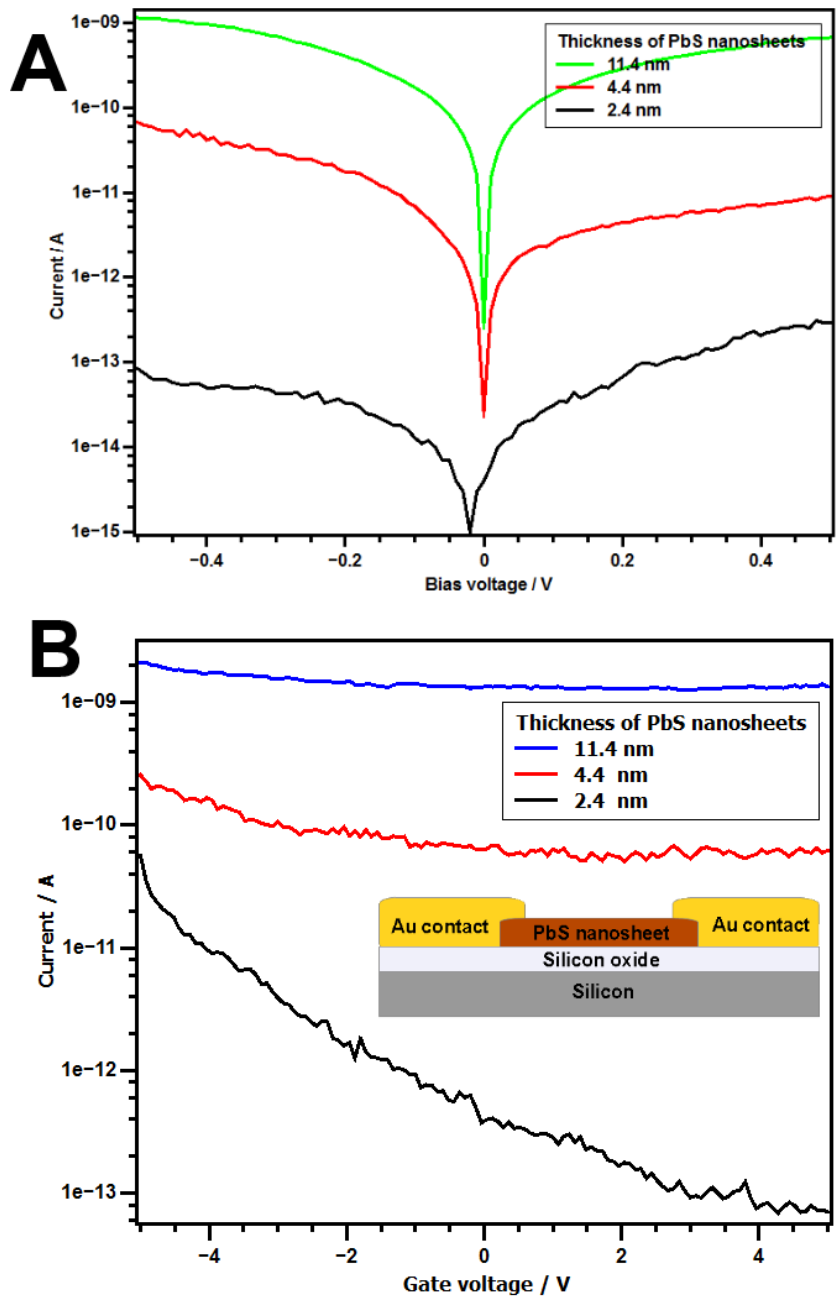

**Figure 6.** Electrical transport measurements on individual nanosheets: (A) output characteristics $I_D$-$V_{DS}$ at $V_g = 0$ V and (B) transfer characteristics $I_D$-$V_g$ at $V_{DS} = -0.5$ V of PbS nanosheets of various thicknesses (2.4 nm, 4.4 nm, and 11.4 nm) at room temperature plotted on a logarithmic scale. The inset in (B) shows a sketch of the devices.